\documentstyle[11pt]{article}

\def\pa{\partial}

\def\a{\alpha}
\def\b{\beta}
\def\d{\delta} 
\def\e{\epsilon}

\def\l{\lambda} 
\def\m{\mu}
\def\n{\nu}

 \def\O{\Omega}
\def\mn{{\mu\nu}}
\def\ab{{\alpha\beta}}
\def\be{\begin{equation}}
\def\ee{\end{equation}}

\def\bea{\begin{eqnarray}}
\def\eea{\end{eqnarray}}

\def\half{\frac{1}{2}}

\begin{document}

\begin{flushright} BRX TH-565 \end{flushright}

\vspace*{.3in}

\begin{center}
{\Large\bf Birkhoff for Lovelock Redux}

{\large S.\ Deser$^1$ and J.\ Franklin$^2$}

{\it $^1$Department of Physics,  Brandeis University\\Waltham, MA 02454, USA\\
{\tt deser@brandeis.edu}

$^2$ MIT Kavli Institute for Astrophysics and Space Research \\  Cambridge, MA 02139, USA\\
{\tt jfrankli@mit.edu}}

\end{center}

\begin{abstract}
 We show succinctly that all metric theories with second order field equations
obey Birkhoff's theorem: their spherically symmetric solutions are static.  \end{abstract}

 
The special, ``Lovelock", metric theories with second derivative order field equations (relevant details, and the original references, may be found in \cite{deserryzhov}, whose notation we use.)
obey Birkhoff's theorem: their spherically symmetric solutions are static \cite{chermousisdufaux,zegers}.  
In this note, we provide an alternative derivation of the theorem, using the approach of \cite{deserfranklin}, 
where historical references-and justification of our shortcuts-are given. We also show that the apparent 
counterexamples to Birkhoff in tuned combinations of Lovelock actions found in \cite{chermousisdufaux,zegers} 
are merely deSitter vacua, such as can always be introduced by tuning any action polynomial in curvatures.

We will focus on the single relevant, (0r), ``radial energy flux" field equation, using the metric ansatz 
 \be
 ds^2 = - a b^2 dt^2 + a^{-1} dr^2 + r^2 d_n\O + 2 fbdrdt
 \ee
 where $(a,b,f)$ depend on $(r,t)$ and $d_n\O$ is the unit
$n$-sphere interval.  As we shall see, the seemingly superfluous $f \sim g_{0r}$ 
component's variation provides precisely the required field equation. 
The Lovelock actions of order $k$ in $D=n+2$ dimensions are defined as
 \be
 I_k = \int d^Dx \: {\cal L}_k \equiv \int d^Dx (-g)^{-1/2} \e^{\m{_1}\ldots}  \e^{\n{_1}\ldots}R_{\m{_1}\m{_2} . .} \ldots   %
R_{\n_{k-1}\n_{k} . .} \; g_{. .}\ldots g_{. .}
 \ee
 There are $k$ curvatures, whose 4$k$ indices contract with the 2D $\e$-indices; the indicated metrics soak up the rest.  
${\cal L}_k$ vanishes for $D < 2k$ and $I_k$ is a topological invariant at $D=2k$. The Palatini, $\d R \sim DD \d g$, and Bianchi, 
$R_{\m [\n\a\b ; \l ]} \equiv 0$, identities guarantee that only the variations of the explicit metrics (but not those of the $R$'s)
survive, leading to field equations ${\cal G}_\mn = 0$ algebraic of order $k$ in the curvatures but without explicit derivatives.  
The leading, $k = 0,1,2$  actions are the cosmological, GR and Gauss--Bonnet terms. The form of every $I_k [a,b]$, upon inserting  
the above metric, first at $f=0$, is \cite{deserryzhov}
 \be
 I_k [a, b] \sim \int dr b [r^{n+1} \psi^k ]^\prime  \hbox{, \hskip .5cm}  r^2\psi = (1- a)
 \ee
where primes denote radial- (and dots will mean time-) derivatives.  Note first that, even though we have {\it not} assumed time 
independence of $(a,b)$, only {\it spatial} derivatives (that there is only one prime, rather than two, is peculiar to Schwarzschild gauge.) survive \cite{desertekin}.
Further, (3) will be unaffected by subsequent inclusion of $f$, since the latter will only be used as a Lagrange multiplier to deliver 
the $(0r)$ equation, then set to zero, to recover the 2-function ($f=0$) metric.  We learn from varying (3) with respect to $a$ that $b=b(t)$, 
which means it can be removed entirely by using the residual time-gauge choice  $b(t)dt = d\tilde{t}$. [Thence our unusual, but legitimate, 
parametrization  of $ds^2$.] Varying with respect to $b$ fixes  $a(r,t)$, {\it e.g.}\ $a=1-2m(t)/r^{n-1}$ in GR; the notation emphasizes 
that $m$'s time-dependence is as yet undetermined.  For that, we need the final, $f$-field equation,
 \be
\left. \frac{\d I_k [a,f]}{\d f} \right|_{f=0} \equiv F_{k-1} (a) \dot a = 0
\, 
 \ee 
where we can use the now known $(a,b)$ to remove $b$ altogether and as explained, $f$ has also been set to 0 after variation. 
We have-legitimately-anticipated the form of this equation, simply because it must have exactly one time derivative and only 
depends on $a(r,t)$.  So  Birkhoff is proved, unless $F_{k-1}(a)=0$, which is certainly impossible for any one $k$ or we would 
have lost an entire field equation! We will show below that this is also the case for all linear combinations, $\sum_k \,c_k \, I_k$.  
So the only work required 
is to justify that (4) follows from (2) at $b=1$, $f=0$. In varying the action to get its $(0r)$ part, we need not keep the explicit 
$(-g)^{-1/2}$, since $\d ln (-g) = g^\mn \d g_\mn$.  The $(D-2k)\; g_\mn$ factors of (2) all vary identically, so we find
 \be
 \left. \frac{\d I_k }{ \d g_{0r} } \equiv {\cal G}^{0r} \sim \e^{0\ldots}    %
\e^{r\ldots} g\ldots g \; R\ldots R\right |_{f =0} \; = 0 .
 \ee
Because $g_{0i} = 0$, the remaining $g_\ab$ must all be spatial (since only one ``0" index is available); then all curvatures 
but one are purely spatial, the remaining one being $\sim R_{0ijk}$.
 
For orientation, let us briefly consider (5) in Riemann normal coordinates: the $R_{ijk\ell}$ are proportional to $a$ and 
its radial derivative, while $R_{0ijk} \sim (\dot{g}_{ij,k} -\dot{g}_{ik,j})$ has a single $\pa /\pa t$.  Hence, we are morally
assured that the ``guess" (4) is correct, and indeed that $F_{k-1}(a)$ is just the $(k-1)^{\rm th}$ power of $F_1 = (1-a )$, 
the Einstein value.  The Appendix confirms this:  all spatial curvature components depend on $(1-a)$ and $(1-a)'$.

Finally, we turn to generic Lovelock combinations whose actions are of the form~\cite{deserryzhov}
\begin{equation} 
I = \sum_{k=0}^N c_k I_k = \sum_{k=0}^N c_k \int dr b (r^{n+1} \psi^k)'  \hbox{, \hskip .5cm}  r^2\psi = (1- a)
\end{equation}
plus terms linear in $f$.  Clearly, $b = b(t)$ again and gets absorbed by a change of $t$ while $\psi$ obeys the 
algebraic equation $\sum c_k \psi^k = m(t) r^{-(n+1)}$.  Finally, as predicted, the ``Birkhoff equation", is
\begin {equation} 
\left(\sum_{k=1}^N d_k \psi^{k-1} \right) \dot\psi  = 0;
\end{equation}
the $d_k$ differ from the original constants $c_k$ by constant numerical factors. As we noted, a cosmological term ($k=0$) 
would not contribute explicitly.  So $\psi$ is time-constant unless the prefactor in (7) vanishes.  
This vanishing was interpreted in~\cite{chermousisdufaux, zegers}, in a different gauge, to imply that there are 
Birkhoff-violating solutions.  We now see instead that the theorem always holds, and that these exceptional geometries are nothing but (anti) deSitter vacua that generalize the original flat one by fine-tuning the $c_k$:  
From (7), the zeros of the polynomial are its (necessarily time-constant) roots, $\psi_i = R_i$, namely the constant curvature metrics $a_i = 1 + R_i r^2$.  These deSitter spaces not only obey Birkhoff, but the effective cosmological constants can be written, upon inserting the $a_i$ into the $a$-field equation, in terms of the $c_k$ coefficients, including (if any) the original $c_0 = \Lambda$.  Their significance as additional vacua is understood as follows:  Suitable choices of the constants 
in any  $I = \sum \gamma_k R^k$, Lovelock or not, can be fine-tuned to allow constant curvature spaces  for any desired $\Lambda$.  These vacua actually agree with the seemingly time-dependent spaces of~\cite{chermousisdufaux, zegers} upon transforming the latter to Schwarzschild gauge, namely choosing the coefficient of $d_n\O$ to be $r^2$. Finally, this also shows that we have not lost a 
field equation here either-it has just ``solved itself" to yield deSitter!

\section*{Appendix}
 
In this Appendix, we first provide the form of the Riemann tensor for our time-dependent metric ansatz; it depends on five functions, reducing to ~\cite{deserryzhov} for $\dot a = \dot b = 0$:  
 
\begin{eqnarray}
R_{\mu\nu\alpha\beta} &=& 4 A \delta^0_{[\mu}\delta^r_{\nu]} \delta^0_{[\alpha} \delta^r_{\beta]}+ 
4 B Z_i \delta^0_{[\mu} \delta^{\theta_i}_{\nu]} \delta^0_{[\alpha} \delta^{\theta_i}_{\beta]} +
4 C Z_i \delta^r_{[\mu} \delta^{\theta_i}_{\nu]} \delta^r_{[\alpha} \delta^{\theta_i}_{\beta]} + 
2 r^4 \psi Z_i Z_j \delta^{\theta_i}_{[\mu} \delta^{\theta_j}_{\nu]} \delta^{\theta_i}_{[\alpha} \delta^{\theta_j}_{\beta]} \nonumber \\
&+ &4 D Z_i \left( \delta^0_{[\mu} \delta^{\theta_i}_{\nu]} \delta^r_{[\alpha} \delta^{\theta_i}_{\beta]} +
 \delta^r_{[\mu} \delta^{\theta_i}_{\nu]} \delta^0_{[\alpha} \delta^{\theta_i}_{\beta]} \right)\label{RiemannDecomp},
\end{eqnarray}
with
\begin{eqnarray}
A &=& \frac{b}{2} \left[ 3 a' b' + a''b + 2  a b'' \right] + \frac{1}{2 a^2} \left[ -\frac{\dot a \dot b}{b} + \frac{a \ddot a - 2 (\dot a)^2}{a} \right] \\
B &=& \half(r a b) \left[a'b + 2 a b'\right] \\
C &=& -\frac{a'r}{2a}, \hbox{\hskip .5cm}
D = -\frac{\dot a r}{2 a}, \hbox{\hskip .5cm}
\psi = \frac{1 - a}{r^2} 
\end{eqnarray}
and $Z_i$ is the $i^{\hbox{th}}$ angular component of the metric.  

We are interested in relating ${\cal G}^{0r}(k)$ and ${\cal G}^{0r}(k-1)$  to exhibit their ``polynomial in $\psi$" nature,  outlining the steps here. We find that effectively, the Riemann tensor appears in the field equation in the combination
\begin{equation}\label{almost}
\tilde R_{\alpha \beta \gamma \rho} = 2 \psi g_{\alpha \beta} g_{\gamma \rho} + 4 D r^{-2} g_{\gamma \rho} \delta^r_{\alpha} \delta^0_\beta.
\end{equation}
The first term above is ${\cal G}^{0r}(k-1)$ while the second is (in odd dimensions) $k^{-1}G^{0r}(k)$ when replacing the $k^{\hbox{th}}$ Riemann tensor in the field equation (5). 
After some algebra, this leads to 
\begin{equation}
{\cal G}^{0r}(k) = 2 \psi {\cal G}^{0r}(k-1) + k^{-1} {\cal G}^{0r}(k) \rightarrow {\cal G}^{0r}(k) = \frac{2 k \psi}{k - 1} {\cal G}^{0r}(k-1),
\end{equation}
and so
\begin{equation}
{\cal G}^{0r}(k) = (\hbox{const.}) \times \psi^{k-1} {\cal G}^{0r}(1);
\end{equation}
the final term here is Einstein, and contains a factor of $D$ and a product of all the angular dependence but no $\psi$.

Our other task here is, for concreteness, to show explicitly in D=5  how the seeming obstructions to Birkhoff are really 
deSitter vacua, as shown generically in text.

For the metric ansatz:
\begin{equation}
ds^2 = -a b^2 dt^2 + a^{-1} dr^2 + 2 b f dt dr + r^2 (d\psi^2 + \sin^2\psi( d\theta^2 + \sin^2\theta d\phi^2))
\end{equation}
we form the cosmological, Einstein, and Gauss-Bonnett actions
\begin{equation}
I = \int d^5 x \sqrt{-g} \left( \Lambda + R + \alpha( R^2 - 4 R^{\mu\nu} R_{\mu\nu} + R^{\mu\nu\alpha\beta} R_{\mu\nu\alpha\beta}) \right),
\end{equation}
keeping only terms linear in $f$ and its derivatives.  The field equations at $f = 0$ read:
\begin{eqnarray}
\frac{\delta I}{\delta a} \biggr\vert_{f = 0} &=& 3 (4 \alpha(1 - a) + r^2) b' = 0 \\
\frac{\delta I}{\delta b} \biggr\vert_{f=0} &=&  \Lambda r^3 + 6 r (1 - a) - 3 r^2 a' - 12 \alpha a' (1 - a) \\
{\cal G}^{0r} &\equiv& \frac{\delta I}{\delta f} \biggr\vert_{f=0} = -3(r^2 +4 \alpha(1 - a) ) \dot a a^{-1}.
\end{eqnarray} 
Now consider the obstruction, $4 \alpha (1 - a) + r^2 = 0$, to Birkhoff. This is just $a = 1 + \frac{r^2}{4 \alpha}$, 
vacuum deSitter. The constant $4 \alpha$ is determined by inserting into the $b$ equation:
\begin{equation}
\frac{\delta I}{\delta b} \biggr \vert_{f = 0, a = 1 + r^2/(4 \alpha)} = \frac{r^3 (2 \alpha \Lambda - 3)}{2 \alpha} = 0.
\end{equation} 
The Lovelock coefficients determine the deSitter cosmological constant according to the tuning $\alpha = 3 \Lambda/2$;  in the absence of $\Lambda$, there are no roots, except flat space.

We thank B.\ Tekin for a discussion. This work was supported by NSF grant PHY-04-01667.

\end{document}